\documentstyle[amstex,12pt]{article}
\begin{document}

\begin{center}
{\bf Interplay of static and dynamic effects in $^{6}$He+$^{238}$U Fusion}\\[%
0pt]
\bigskip W.H.Z. C\'{a}rdenas$^{1}$, L.F.~Canto$^{2}$, N.~Carlin$^{1}$,
R.~Donangelo$^{2}$\ and M.S.~Hussein$^{1}$\\[0pt]
$^{1}$Instituto de F\'{\i}sica, Universidade de S\~{a}o Paulo,\\[0pt]
C.P. 66318, 05389-970 S\~{a}o Paulo, Brazil\\[0pt]
$^{2}$Instituto de F\'{\i}sica, Universidade Federal do Rio de Janeiro, \\[%
0pt]
C.P. 68528, 21941-972 Rio de Janeiro, Brazil\\[0pt]
\end{center}

\bigskip

\noindent {\bf Abstract}

\noindent We investigate the influence of the neutron halo and the breakup
channel in $^{6}$He + $^{238}$U fusion at near-barrier energies. To include
static effects of the 2n-halo in $^{6}$He nuclei, we use a single-folding
potential obtained from an appropriate nucleon-$^{238}$U interaction and a
realistic $^{6}$He density. Dynamical effects arising from the breakup
process are then included through coupled-channel calculations. These
calculations suggest that static effects dominate the cross section at
energies above the Coulomb barrier, while the sub-barrier fusion cross
section appears to be determined by coupling to the breakup channel. This
last conclusion is uncertain due to the procedure employed to measure the
fusion cross-section. \bigskip

\section{Introduction}

The recent availability of radioactive beams has made possible to study
reactions involving unstable nuclei \cite{BCH93}. Several of the light
neutron and proton rich nuclei exhibit halo structures, with a compact core
plus one or two loosely bound nucleons. For example, $^{11}$Li and $^{6}$He
are two-neutron, borromean halo nuclei, while $^{11}$Be and $^{19}$C are
one-neutron halo nuclei. The isotope $^{8}$B has been confirmed to have a
one-proton halo, while $^{17}$F is a normal nucleus in its ground state but
becomes a one-proton halo in its first excited state.

Reactions induced by these nuclei are important in processes of
astrophysical interest, among others. We ask the question of how the above
systems fuse, in particular how the fusion induced by these nuclear species
behaves as a function of bombarding energy, especially near the Coulomb
barrier.

The main new ingredient in reactions induced by unstable projectiles is the
strong influence of the breakup channel. One important feature of these
loosely bound systems is that they exhibit the so-called soft giant
resonances (pygmy resonances), the most notorious of which is the soft
dipole resonance, very nicely confirmed in $^{6}$He by Nakayama {\it et al.} 
\cite{Naka02}.

In the case of not too unstable projectiles, the effect of the breakup
channel in the fusion cross section at sub-barrier energies is, as in the
case of stable beams, to enhance it. At slightly above-barrier energies,
however, the situation is qualitatively different from the case where only
stable nuclei are involved. The contribution from the breakup channel to the
fusion reaction is strongly influenced by the low probability that all
fragments are captured. Thus, in this case, the fusion cross section is
partitioned into a complete and one or more incomplete fusion contributions.

Recently, nuclear reactions involving the neutron-rich nucleus $^{6}$He have
attracted considerable attention \cite{Au99}. In particular, very interesting
experimental data on the fusion of He isotopes with $^{238}$U have been
obtained \cite{Tr00}. These data show an enhancement of several orders of
magnitude of the $^{6}$He+$^{238}$U fusion cross section with respect to
that of $^{4}$He+$^{238}$U. The physical process leading to this result has
not yet been established. A natural candidate is the coupling with the
breakup channel. This led us to develop a simple theoretical model to
estimate statical and dynamical effects of the breakup channel on the
complete and incomplete fusion cross section in the $^{6}$He+$^{238}$U\
collision. The extension of the model to study fusion induced by other
radioactive beams is straightforward.

The paper is organized as follows. Section 2 describes the
calculation of the static effects brought about by the presence of a nuclear
halo. The coupling to the breakup channel is performed, by means of
schematic coupled-channels calculations, in section 3. Our conclusions are
presented in the last section.

\section{Static effects from the 2n-halo}

The weakly bound neutrons in $^{6}$He are expected to influence the fusion
cross section in two ways. Firstly, by the static effect of barrier lowering
due to the existence of a halo. Secondly through the coupling with the
breakup channel. In this section we consider the first of these effects.

Owing to the two weakly bound neutrons in $^{6}$He, the nuclear density has
a long-range tail and so does the real part of the optical potential
describing the $^{6}$He-target collision. In this way, the potential barrier
is lowered and the fusion cross section is enhanced. In order to account for
this effect, we use a single folding model do describe the real part of the
nuclear $^{6}$He-$^{238}$U interaction. This potential is given by the
expression%
\begin{equation}
V_{N}({\bf r})=\int v_{n-T}({\bf r}-{\bf r}^{\prime })\,\rho ({\bf r}%
^{\prime })\,d^{3}{\bf r}^{\prime }\,.  \label{sf}
\end{equation}%
Above, $v_{n-T}({\bf r}-{\bf r}^{\prime })$ is a phenomenological
interaction between a nucleon and the $^{238}$U target nucleus and $\rho (%
{\bf r}^{\prime })$\ is a realistic $^{6}$He\ density, containing the
contribution from the halo. The nucleon-$^{238}$U interaction is obtained
from studies of the collision of low-energy neutrons with heavy target
nuclei in the actinide region. It can be written (discarding the spin-orbit
part) \cite{n-238U} 
\begin{equation}
v_{n-T}\,(x)=-V_{0}\,f_{r}(x),  \label{V_n-T}
\end{equation}%
with 
\begin{equation}
V_{0}=\left[ 50.378-27.073\left( \frac{N-Z}{A}\right) -0.354\,E_{Lab}\right]
\,(MeV)
\end{equation}%
and%
\begin{equation}
f_{r}(x)=\frac{1}{1+\exp \left[ \left( x-R_{r}\right) /a_{r}\right] }\,,
\label{W-S}
\end{equation}%
with the parameters $R_{r}=1.264\,A_{T}^{1/3}$ fm and $a_{r}=0.612$ fm. The
total optical potential is then given by

\begin{equation}
U(r)=V_{N}(r)+V_{C}(r)-iW(r)\ .  \label{pmo}
\end{equation}%
Above, $V_{C}(r)$ is the usual Coulomb interaction in nuclear collisions,%
\begin{eqnarray}
V_{C}(r) &=&\frac{Z_{P}Z_{T}e^{2}}{r}\quad \quad \quad \quad \quad \quad
\quad r\geq R_{C}=1.2\,(A_{T}^{1/3}+A_{P}^{1/3})\,\   \nonumber \\
&=&\frac{Z_{P}Z_{T}e^{2}}{2R_{C}}\,\left( 3-\frac{r^{2}}{R_{C}^{2}}\right)
\quad \quad r<R_{C},\quad \quad \quad \quad \quad \quad \quad \quad
\end{eqnarray}%
and $W(r)$\ is a volumetric strong absorption potential with small values
for both its radius and diffusivity. We use the parametrization%
\begin{equation}
W(r)=W_{0}\,\,f_{i}(r),
\end{equation}%
with $W_{0}=50$ MeV and $f_{i}(r)$ a Wood-Saxon shape as in eq.(\ref{W-S})
with 
\begin{equation}
R_{i}=1.0\,\left( A_{P}^{1/3}+A_{T}^{1/3}\right) ~{\rm fm},\quad a_{i}=0.10~%
{\rm fm.}
\end{equation}

As a test, we applied the above procedure to $^{4}$He + $^{238}$U fusion.
The nuclear potential was evaluated by eq.(\ref{sf})\ using a Gaussian
density. We write 
\begin{equation}
\rho (r)=C~\exp \left( -r^{2}/\gamma ^{2}\right) ~  \label{gauss}
\end{equation}%
and choose the parameters $C$ and $\gamma $\ as to give the correct
normalization and experimental r.m.s. radius.\ That is

\begin{equation}
\int \rho (r)d^{3}r=A;\;\int r^{2}~\rho (r)d^{3}r=r_{rms}^{2}~.
\label{param}
\end{equation}%
In the present case, we set $A=4$ and $r_{rms}=1.49$ fm \cite{Al98}. The
fusion cross section obtained with our optical model calculation with the
single folding potential is shown in figure 1 (thin solid line), in
comparison with the data of Trotta {\it et al.} \cite{Tr00}\ and the data of
Viola and Sikkland \cite{VS62}. The agreement is very good. Since the
calculation contains no free parameter, this agreement indicates that the
procedure is reasonable.

We now consider $^{6}$He + $^{238}$U fusion. Firstly, we disregard the
existence of the $^{6}$He halo and repeat the above procedure. We
parametrize the density as in eq.(\ref{gauss}) and scale the density and
r.m.s. radius to $^{6}$He. That is, we set in eq.(\ref{param}) $A=6$ and $%
r_{rms}=\left( 6/4\right) ^{1/3}\times 1.49$ fm. This density is then used
in eq.(\ref{sf}) and the folding potential is determined. The fusion cross
section calculated with this potential is shown in figure 2 (dashed line),
in comparison with the data \cite{Tr00}. The agreement is poor throughout
the considered energy range. We now take into account the existence of the $%
^{6}$He halo, replacing the gaussian of eq.(\ref{gauss}) by a realistic
parametrization \cite{Al98} of the $^{6}$He\ density, based on the
symmetrized Fermi distribution of ref. \cite{Yu73}. It leads to the r.m.s.
radius $r_{rms}=2.30\,$fm. Using this density in eq.(\ref{sf}), we obtain a
potential which includes contributions from the $^{4}$He-core and also from
the 2n-halo. The resulting fusion cross section is represented by a solid
line in figure 2. We note that the agreement with the data at above barrier
energies $\left( E_{c.m.}>V_{B}\simeq 22.3\text{ MeV}\right) $\ is
considerably improved. Since the Coulomb barrier height is reduced by the
attractive contribution from the halo, the cross section becomes larger.
However, at sub-barrier energies the agreement remains very poor. The
theoretical prediction for the fusion cross section is still several orders
of magnitude smaller than the experimental data.

\section{Coupled channel effects}

It is well known that the coupling between channels enhances the fusion
cross section at sub-barrier energies \cite{DLW85}. Therefore,
coupled-channel effects should be taken into account in a theoretical
description of the fusion process. However, in the case of coupling to the
breakup channel the situation is more complicated since the breakup channel
involves an infinite number of continuum states. A possible treatment of the
problem, used in refs. \cite{Ha00,Di02}, is to use continuum discretization
to reduce it to a finite number of channels. The situation is still more
complicated in the breakup of $^{6}{\rm He},$ since it breaks up into three
particles. In the present work we schematically replace the breakup channel
by an {\it effective channel} \cite{Br95}. This state has energy equal to
the breakup threshold and carries the full strength of the continnum. This
procedure is justified when breakup occurs through a low-lying long-lived
resonance (with a half life much larger than the collision time), as it seems
to be the case with $^{6}{\rm He}$ \cite{Naka02}. Since the kinetic energy
of the relative motion between the $^{4}{\rm He}$-core and the neutron pair
is neglected, this approximation tends to overestimate the importance of the
coupling to the breakup channel. Therefore the simplified model of the
present work should provide an upper limit for the fusion cross section.

The starting point of the coupled channel method is the Schr\"{o}dinger
equation for the colliding system,%
\begin{equation}
H\Psi ({\bf r},\xi )=E\Psi ({\bf r},\xi ),  \label{schor}
\end{equation}%
where ${\bf r}$ is the projectile-target vector, $\xi \ $stand for the
relevant intrinsic coordinates, $E$ is the total energy in the center de
mass frame and $H$ is the total Hamiltonian of the system . One then
performs the channel expansion of the wave function 
\begin{equation}
\Psi ({\bf r},\xi )=\sum_{\alpha }\psi _{\alpha }({\bf r})~\phi _{\alpha
}(\xi ),  \label{expa}
\end{equation}%
where $\phi _{\alpha }(\xi )$\ denotes an intrinsic state with energy $%
\epsilon _{\alpha }$\ and$\ \psi _{\alpha }({\bf r})$ is the relative motion
wave function in channel-$\alpha $. Substituting this expansion in eq.(\ref%
{schor}), we obtain the coupled-channel equations 
\begin{equation}
(E_{\alpha }-H_{\alpha })\,\psi _{\alpha }({\bf r})=\sum_{\beta }{\cal V}%
_{\alpha \beta }({\bf r})\,\psi _{\beta }({\bf r}).  \label{coup}
\end{equation}%
Above, $E_{\alpha }$ = $E-\epsilon _{\alpha }$ and $H_{\alpha }$ = $%
T+U_{\alpha }(r)$, where $U_{\alpha }(r)$ is the optical potential in
channel-$\alpha .$ The channels are coupled through an interaction ${\cal V}(%
{\bf r,}\xi ),$\ with matrix-elements in channel space given by%
\begin{equation}
\nu _{\alpha \beta }({\bf r})=\int d\xi ~\phi _{\alpha }^{\ast }(\xi )~~v(%
{\bf r},\xi )~~\phi _{\beta }(\xi )~.  \label{matrix-elements}
\end{equation}%
For practical purposes, it is convenient to carry out angular momentum
expansions. The wave function is then written as (see e.g. \cite{Sa83}) 
\begin{eqnarray}
\Psi ^{(+)}(\alpha _{0}\nu _{0}{\bf k}_{0};{\bf r}) &=&\frac{1}{\left( 2\pi
\right) ^{3/2}}\sum_{JMl_{0}}4\pi \,\,\langle JM\left\vert l_{0}(M-\nu
_{0})\,I_{0}\nu _{0}\right\rangle \,Y_{l_{0}\,(M-\nu _{0})}^{\ast }({\bf 
\hat{k}}_{0})\,  \nonumber \\
&&\times \,\sum_{\alpha l}{\cal Y}_{\alpha l}^{\pi JM}(\zeta )\,\frac{%
u_{\alpha l,\alpha _{0}l_{0}}^{J}(k_{\alpha },r)}{k_{0}r}\,
\label{8Psi-exp-spin}
\end{eqnarray}%
and using this expansion in eq.(\ref{schor}) one obtains the angular
momentum projected coupled channel equations%
\begin{multline}
{\Huge [}E_{\alpha }+\frac{\hbar ^{2}}{2\mu }{\Huge (}\frac{d^{2}}{dr^{2}}-%
\frac{l\left( l+1\right) }{r^{2}}{\Huge )}-{\cal V}_{\alpha l}^{J}(r){\Huge ]%
}\,u_{\alpha l,0l_{0}}^{J}(k_{\alpha },r)  \label{8CC-2} \\
=\sum_{\alpha ^{\prime }l^{\prime }}{\cal V}_{\alpha l,\alpha ^{\prime
}l^{\prime }}^{J}(r)\,\,u_{\alpha ^{\prime }l^{\prime },0l_{0}}(k_{\alpha
^{\prime }},r)\ .  \nonumber
\end{multline}%
In the present calculation, $\alpha $ takes only the values $0$ (elastic
channel) and 1 (effective breakup channel). For the energy of the breakup
channel we used $\epsilon _{1}=0.975$ MeV, which corresponds to the breakup
energy. As said above, this means we neglect the kinetic energy of the
relative motion of the fragments after breakup.

We initially consider the coupling interaction as the electric dipole term
in the multipole expansion of the electromagnetic interaction between the
projectile and the target. This is based on the idea that, in order to break
a very weakly bound nucleus, only a small perturbation is needed. The fact
that the breakup cross section for those nuclei is very large, suggests that
this process is important.

In the case of a electric dipole interaction, the coupling matrix elements
are \cite{Sa83}%
\begin{equation}
{\cal V}_{1l,0l_{0}}^{J}(r)=A~\,i^{l-l_{0}}\,\hat{l}\,\hat{l}_{0}\sqrt{\frac{%
4\pi }{3}}\,\frac{1}{r^{2}}~\left( 
\begin{array}{ccc}
l\, & 1 & l\,_{0} \\ 
0 & 0 & 0%
\end{array}%
\right) \,\left\{ 
\begin{array}{ccc}
J & 1 & l \\ 
1 & l_{0} & 0%
\end{array}%
\right\} ~,  \label{8Voff-3}
\end{equation}%
with%
\begin{equation}
A=eZ_{T}\,\sqrt{B(E1,0\rightarrow 1)}\,\,(-)^{J+1}  \label{AAA}
\end{equation}%
Above, $\left( 
\begin{array}{ccc}
l\, & 1 & l\,_{0} \\ 
0 & 0 & 0%
\end{array}%
\right) $ and $\left\{ 
\begin{array}{ccc}
J & 1 & l \\ 
1 & l_{0} & 0%
\end{array}%
\right\} $ are the usual 3J and 6J symbols \cite{Ed74}. Note that the above
matrix-elements are fully determined, except for the value of the reduced
transition probability $B(E1,0\rightarrow 1).$

Solving the coupled channel equations, one obtains the fusion cross section
by the formula\footnote{%
The constant $\left( 2\pi \right)^{3}$ in the expression for $\sigma_{F}$
arises from the normalization factor $\left( 2\pi \right)^{-3/2}$ adopted
for $\psi _{\alpha }^{(+)}.$} 
\begin{equation}
\sigma _{F}=\left( 2\pi \right) ^{3}~\frac{k_{0}}{E}\,\sum_{\alpha =0}^{1}%
{\bf \,}\left\langle \psi _{\alpha }^{(+)}\right\vert W_{\alpha }\left\vert
\psi _{\alpha }^{(+)}\right\rangle \,.  \label{8sig_a-total}
\end{equation}

The method of the present work was used to evaluate the fusion cross section
in the $^{6}$He+$^{238}$U collision. We used the optical potential discussed
in the previous section, which include the static effects of the halo. The
coupling matrix-elements were given by eqs.(\ref{8Voff-3}) and (\ref{AAA}),
with the $B(E1,0\rightarrow 1)$ given by the cluster model \cite{BCH93}, 
\begin{equation}
B(E1,0\rightarrow 1)=\frac{3\hbar ^{2}e^{2}}{16\pi \epsilon _{1}\mu _{2n-^{4}%
{\rm He}}}~.  \label{BE1-cluster}
\end{equation}%
Above, $\epsilon _{1}\ $is the energy binding the {\it dineutron} to $^{4}%
{\rm He}$ in the $^{6}{\rm He}$ nucleus and $\mu _{2n-^{4}{\rm He}}$\ is the
corresponding reduced mass.\ Taking the numerical value of eq.(\ref%
{BE1-cluster}), we obtain $B(E1,0\rightarrow 1)=1.37\,~e^{2}$ fm$^{2}$.

Recently Hagino {\it et al.} \cite{Ha00} have shown that the effects of the
nuclear coupling may extend quite far in the case of weakly bound nuclei. In
order to estimate the additional dynamic effects arising from the nuclear
interaction, we must include the coupling due to the nuclear potential.
Since we use an effective channel to describe breakup states, the
calculation of the nuclear form factor is a complicated task. For the
estimates of the present work, we considered the nuclear interaction
potential associated to $^{6}{\rm He}$ breakup to be the difference between
the sum of the nuclear potentials between $^{238}$U and $^{4}{\rm He}$ and
the dineutron, and between $^{238}$U and the $^{6}{\rm He}$ projectile, {\it %
i.e.} 
\begin{equation}
V_{int}^{N}({\bf r},{\bf x})=V_{^{4}{\rm He}}({\bf r+x/}3)+V_{2n}({\bf r-}2%
{\bf x/}3)-V_{^{6}{\rm He}}({\bf r})\,.
\end{equation}%
Above, ${\bf x}$ is the vector going from the di-neutron to the $^{4}{\rm He}
$ cluster, $V_{2n}$ is twice the potential of eq.(\ref{V_n-T}) and $V_{^{4}%
{\rm He}},\,$and $V_{^{6}{\rm He}}$ are the folding potentials of the
previous section. We carry out the angular momentum expansion%
\begin{equation}
V_{int}^{N}=\sum_{\lambda ,\mu }Y_{\lambda \mu }\left( {\hat{r}}\right)
Y_{\lambda \mu }^{\ast }\left( {\hat{x}}\right) ~V_{\lambda }^{N}(r,x)
\end{equation}%
and keep only the dipole term ($\lambda =1$). In this way, the nuclear form
factor is%
\begin{equation}
F_{\lambda =1}^{N}(k;r)=\int_{0}^{\infty }dr\,\phi
_{0}(x)~V_{1}^{N}(r,x)~u_{1}(k,x)\,,  \label{FN}
\end{equation}%
where $\phi _{0}(x)$ is the radial function associated to the bound state of
the 2n-$^{4}$He system and $u_{1}(k,x)$ the $l=1$ continuum radial
wavefunction for the same system, with energy $E_{k}=\hbar ^{2}k^{2}/2\mu
_{2n-^{4}{\rm He}}$. Both functions are calculated using the radial Schr\"{o}%
dinger equation associated to the internal coordinate ${\bf x}$. The depth
of the $V_{2n-^{4}{\rm He}}$ potential was set in order to have the second
S-state with energy $E_{0}=-0.975$ MeV\ (to be consistent with Pauli
Principle we discarded the first S-state). Owing to the normalization of $%
u_{1}(k,x),$ the above form factor vanishes in the $k\rightarrow \infty $
limit. However, the absolute strength of $F_{\lambda =1}^{N}$ should be
treated as a free parameter, since the final state is an effective channel.
In this way, we adopt the form factor%
\begin{equation}
F_{1}^{N}(r)=F_{0}~f(r)~,  \label{FF1}
\end{equation}%
with%
\begin{equation}
f(r)=\lim_{k\rightarrow \infty }\left[ \frac{F_{\lambda =1}^{N}(k;r)}{%
F_{\lambda =1}^{N}(k;0)}\right] ~.  \label{FF}
\end{equation}%
To estimate the strength $F_{0},$ we adopt the following procedure. Firstly,
we evaluate the Coulomb form factor as we evaluated the nuclear one. Instead
of using $B(E1,0\rightarrow 1)=0.59\,~e^{2}$ fm$^{2},$ we calculate reduced
matrix elements of the dipole term in the Coulomb coupling using the analog
of eq.(\ref{FN}). The resulting Coulomb and nuclear dipole form factors are
shown in figure 3. Since the dipole term of the nuclear coupling cannot be
written as a product of a function of $r$\ times a function of $x,$\ as can
the Coulomb coupling to a good approximation, the shape of the nuclear form
factor depends on the energy of the continuum state in the $x$-space.
However, the shape of the nuclear form factor does not change much as $%
k\rightarrow 0.$ Although both form factors go to zero in this limit, they
decrease by a common factor. In figure 3, we show the Coulomb and the
nuclear form factors for a very low energy in the continuum. We see that the
ratio of these form factors changes strongly with the radial distance. The
Coulomb form factor dominates at large separations while the nuclear form
factor is larger at small separations. They have approximately the same
strength at $r\simeq 16$ fm. In the present calculation, we use the
experimental $B(E1,0\rightarrow 1)$ value and choose the parameter $F_{0}$
such that the ratio between the nuclear and the Coulomb form factors is
maintained.

Figure 4 shows the $^{6}$He + $^{238}$U total fusion data in comparison to
the static (dashed line) calculation of the previous section, and two
coupled channels calculations. The thin line is the coupled channel
calculation restricted to Coulomb breakup. We notice that the cross section
at high energies is little affected by the inclusion of the breakup channel.
Although the sub-barrier cross section is larger than that found in the
previous section, it remains much smaller than the experimental values.

The solid line is the calculation including also the nuclear coupling. We
notice that it also changes little the cross section at high energies, and
although the nuclear coupling affects more the fusion cross section at
sub-barrier energies, the slope remains much larger than that suggested by
the data. Changing the strength or diffuseness parameters of this coupling
does not change this behavior.

It should be pointed out that the coupling with excited states of $^{238}$U
is not likely to be relevant for this issue, since they were not necessary
for the description of the $^{4}$He+$^{238}$U fusion data, considered in
section 2. As our calculation should provide an upper limit for the cross
section, the experimental fusion cross section at the lowest energies cannot
be explained through our calculations. However, one should keep in mind that
in the calculations presented here we have not included effects due to
coupling to other channels other than breakup, and in particular the
transfer channels. As transfer close to the optimal Q-value may be quite
important at sub-barrier energies \cite{Sw74}, coupling to those channels,
which should not affect much the $^{4}$He+$^{238}$U fusion, is expected to
influence strongly sub-barrier $^{6}$He+$^{238}$U fusion. This could also be
the case for the $^{6}$He+$^{209}$Bi total fusion cross section, where the 
data \cite{Ko98} show a similar trend as sub-barrier energies.

\section{Conclusions}

We have investigated static and dynamic effects on the $^{6}$He+$^{238}$U
fusion cross section. Static effects of the halo were taken into account
through the use of an appropriate optical potential. This potential was
obtained by the single folding model, with a nucleon-target interaction
which is able to reproduce the $^{4}$He+$^{238}$U fusion cross section and
from a realistic $^{6}$He density. Dynamical effects were considered in a
simplified coupled channel calculation, in which the breakup channel was
represented by a single state with energy $\epsilon _{1}=0.975$ MeV (the
threshold for $^{6}$He breakup), concentrating all the low energy dipole
strength. From our calculations we concluded that the static effects
dominate the behavior of the fusion cross section at energies above the
Coulomb barrier. The dynamic coupling to the breakup channel is important
mostly below the barrier. It may be separated into the Coulomb and nuclear
contributions. Although the breakup process takes place at large distances,
we have shown that the coupling with the breakup channel cannot reproduce
the main trends of the data in the sub-barrier region. We believe that
demonstrates that a full description of the $^{6}$He+$^{238}$U fusion cross
section at sub-barrier energies requires the inclusion of neutron-transfer
channels. We point out that a similar enhancement of the sub-barrier fusion
cross section has also been observed in the collision of $^{6}$He with
$^{209}$Bi.

After the completion of this paper we have learned \cite{Tr03} that the data
of Trotta {\it et al}. \cite{Tr00}\ have been reanalized and new data with a
different experimental set up have been taken. The new set of data seems to
indicate that the large enhancement at sub-barrier energies is due to
transfer-fission, rather than fusion-fission events. This is consistent
with our previous remarks.

\bigskip 

\begin{description}
\item[Acknowledgment] The authors are grateful to Prof. H. D. Marta, from
the Instituto de F\'{\i}sica, Facultad de Ingenier\'{\i}a, Montevideo,
Uruguay, for his help in the calculation of the nuclear form factors.
We thank Drs. Christian Beck and Monica Trotta for communicating the features
of their new experimental results prior to their publication. 
This work was supported in part by CNPq and the MCT/FINEP/CNPq(PRONEX) under
contract no. 41.96.0886.00. L.F.C. acknowledges partial support from the
FAPERJ, and M.S.H. and W.H.Z.C. acknowledge support from the FAPESP.\medskip
\end{description}

\noindent {\bf Figure captions} \bigskip

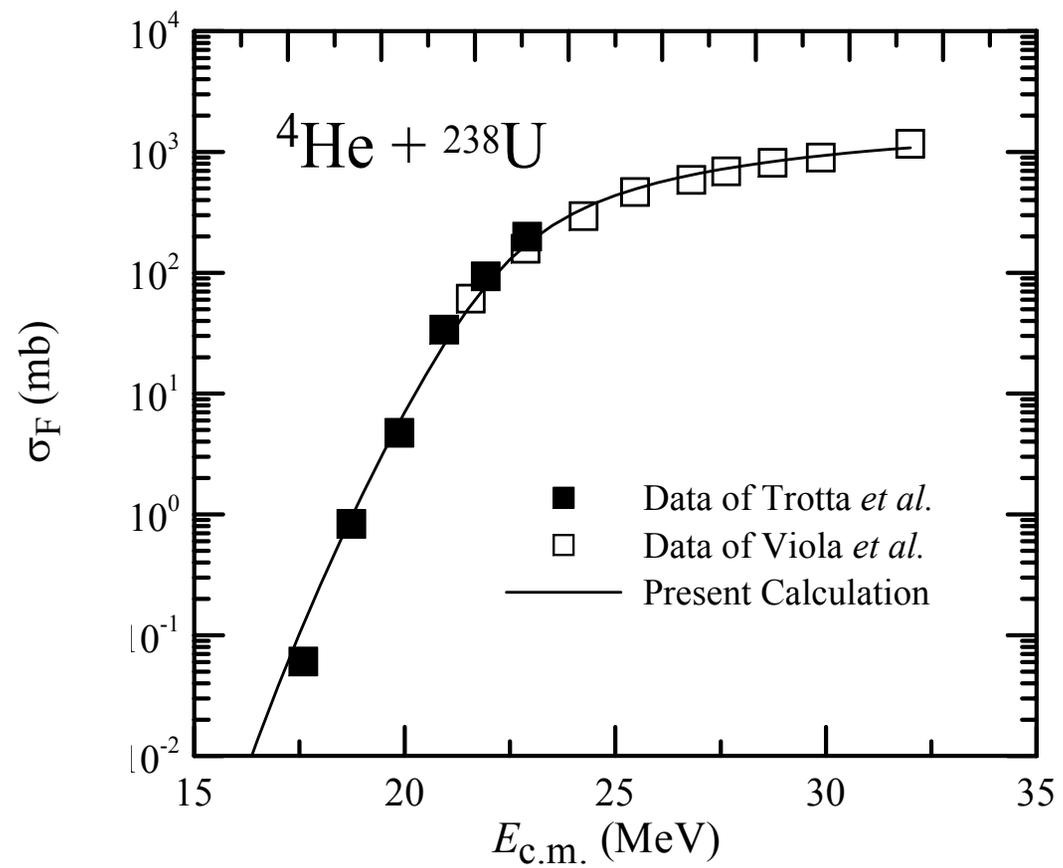
\noindent Figure 1: $\ ^{4}$He+ $^{238}$U fusion cross sections. The data of
refs.\cite{Tr00} (solid squares) and ref.\cite{VS62}\ (open squares) are
compared with the calculations of the present work. The barrier energy is
indicated by an arrow. For further details see the text.

\bigskip

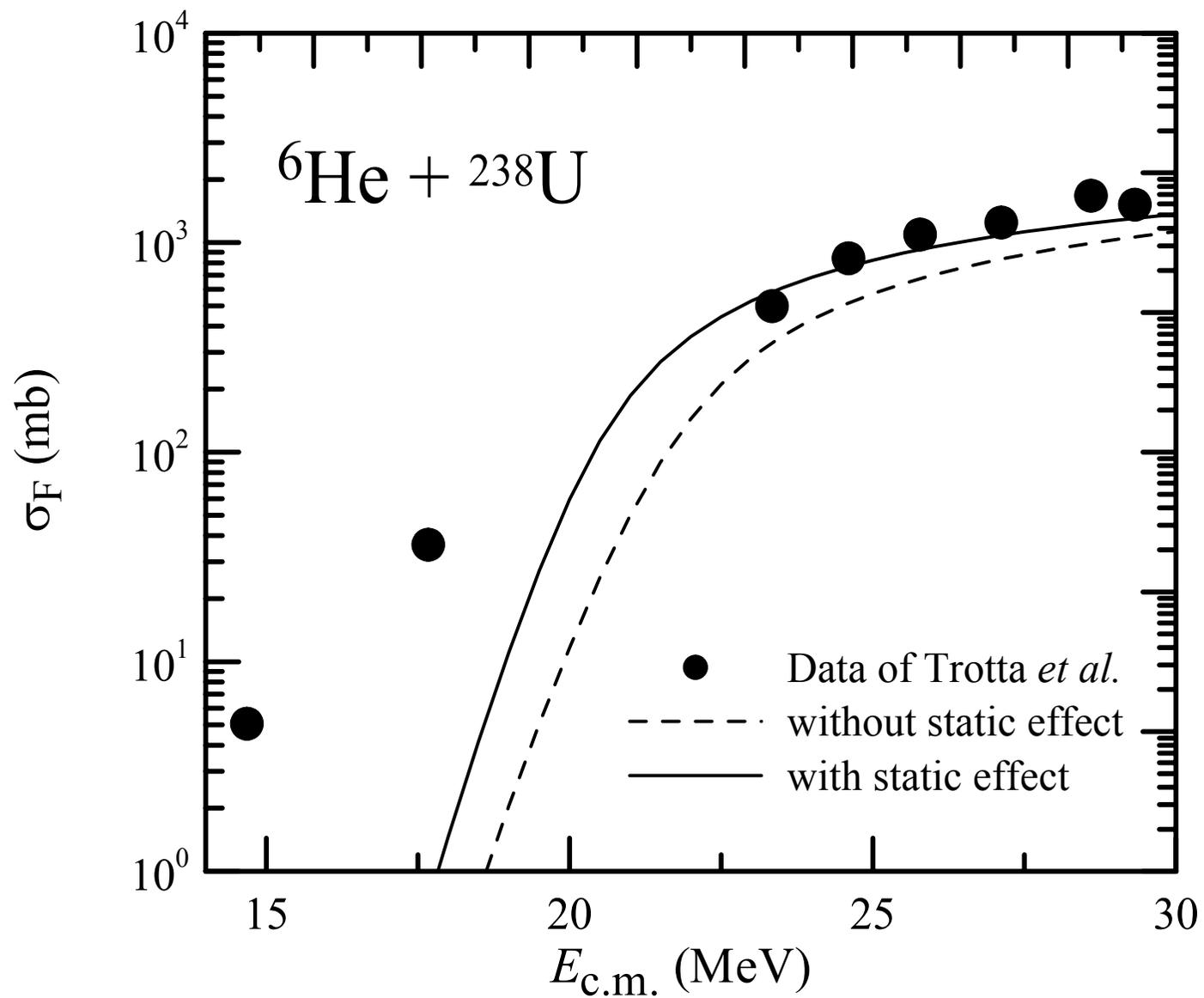
\noindent Figure 2: Coulomb coupling to the breakup channel for the $^{6}$He
+ $^{238}$U fusion cross section. Experimental results \cite{Tr00} are
compared with a static calculation similar to that of figure 1, with just a
scaling of the potential $^{4}$He (dashed line), and taking into account the
fact that $^{6}$He is a halo nucleus (full line).

\bigskip

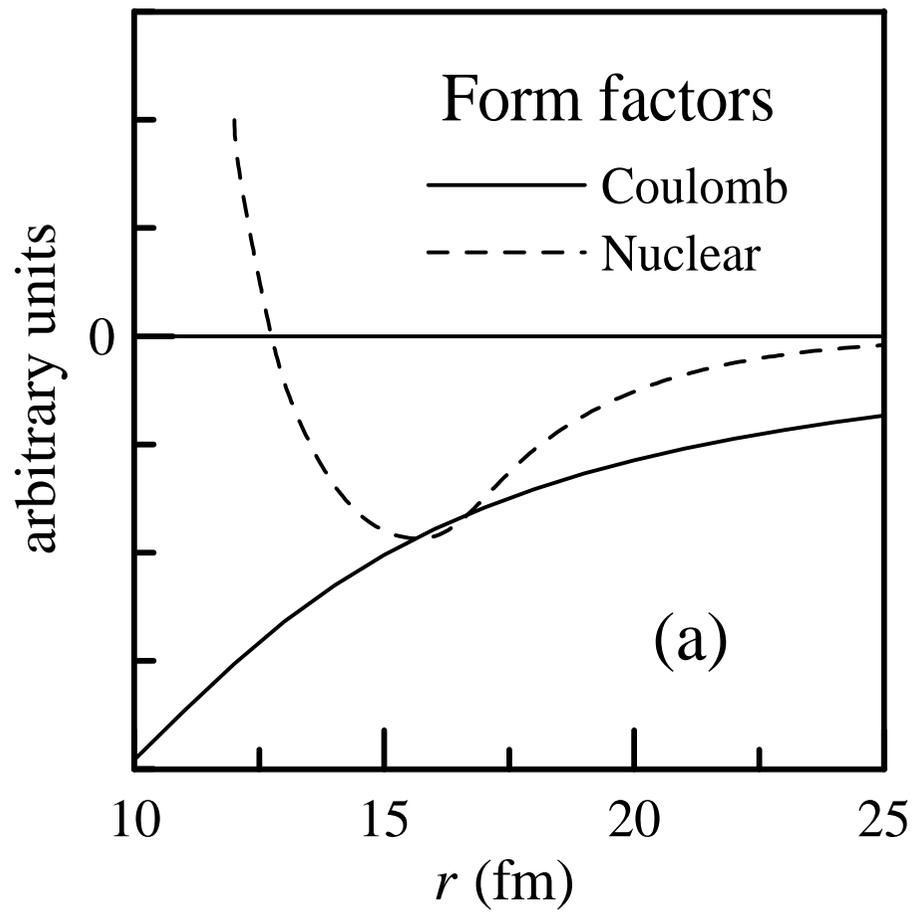
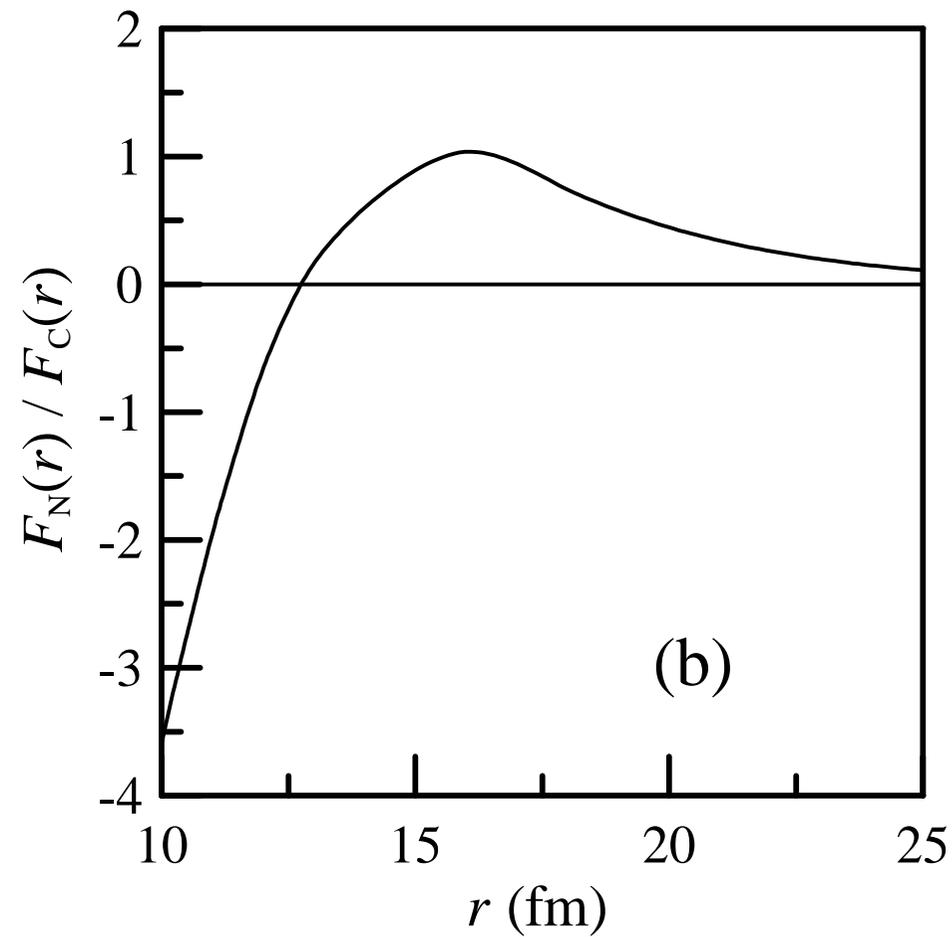
\noindent Figure 3: Coulomb and nuclear dipole form factors (a) and their
ratio (b). See text for details.

\bigskip

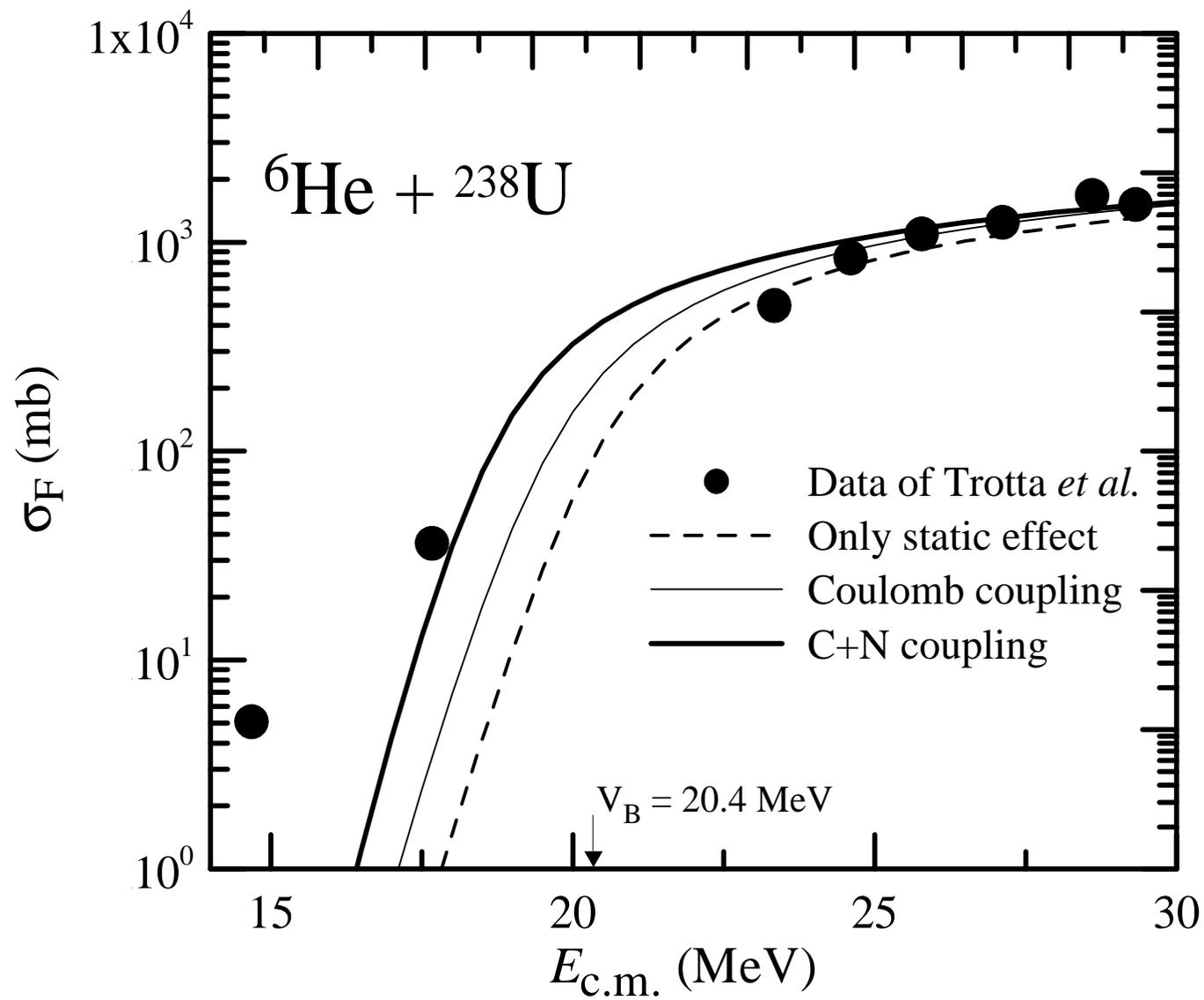
\noindent Figure 4: Total fusion data of $^{6}$He incident on $^{238}$U in
comparison to the static calculation, including the $^{6}$He halo, of figure
2 (dashed line), a coupled channel calculation including only the Coulomb
interaction (thin full line), and also including nuclear effects (thick full
line). See text for details on these two last calculations.

\end{document}